\documentstyle[prb,aps]{revtex}
\draft
\begin{document}

\title{\large \bf {Wavefunctions for the Luttinger liquid}}
\author{K.-V. Pham \and M. Gabay \and P. Lederer}

\address{Laboratoire de Physique des Solides\\ associ\'{e} au CNRS\\
Universit\'{e} Paris--Sud\\ 91405 Orsay, France}

\maketitle
\begin {abstract}
{Standard bosonization techniques lead to phonon-like excitations in a
Luttinger liquid (LL), reflecting the absence of Landau quasiparticles in
these systems. Yet in addition to the above excitations some LL are known to
possess solitonic states carrying fractional quantum numbers (e.g. the spin $%
1/2$ Heisenberg chain). We have reconsidered the zero modes in the
low-energy spectrum of the gaussian boson LL hamiltonian both for fermionic
and bosonic LL: in the spinless case we find that two elementary excitations
carrying fractional quantum numbers allow to generate all the charge and
current excited states of the LL. We explicitly compute the wavefunctions of
these two objects and show that one of them can be identified with the 1D
version of the Laughlin quasiparticle introduced in the context of the
Fractional Quantum Hall effect. For bosons, the other quasiparticle
corresponds to a spinon excitation. The eigenfunctions of Wen's chiral LL
hamiltonian are also derived: they are quite simply the one dimensional
restrictions of the 2D bulk Laughlin wavefunctions \cite{note} .}
\end{abstract}
\pacs {Pacs Numbers: 71.10.Pm 71.27.+a}

 Many one dimensional gapless
quantum systems admit a low-energy effective description similar to that of
the Tomonaga-Luttinger model. This led Haldane to propose the phenomenology
of the Luttinger liquid (LL) to describe these models\cite{haldane}.
Luttinger liquids are quantum critical theories and in the language of
conformal field theories (CFT) they are said to belong to the $c=1$
universality class\cite{ginsparg}. The gaussian boson hamiltonian is a
member of that class, which helps explain why the LL phenomenology is in
essence that of an harmonic acoustic hamiltonian. An important property of
the LL is the absence of Landau quasiparticles\cite{voit}: it is therefore
often believed that the only relevant excitations in a LL are phonons. Yet
let us consider the Heisenberg spin chain which is a well known LL\cite
{luther}; from Bethe ansatz one finds indeed that $\Delta S=1$ excitations
are not magnons, in line with the expectation that there are no Landau
quasiparticles in a LL. In fact $\Delta S=1$ excitations consist of two spin
one-half spinons forming a continuum\cite{fadeev}. It is important to stress
that spinons are fractional excitations of a spin chain (i.e. carrying
fractional quantum numbers) since $\Delta S$ must be an integer for
physically allowed excitations: this is quite clear if we perform
Jordan-Wigner or Holstein-Primakov transformations on the spin hamiltonian
because then a spin $\Delta S=1/2$ excitation becomes a charge $Q=1/2$
excitation in a model for fermions or bosons. Yet it is noteworthy that
there exists no first principle derivation of fractional excitations within
the framework of the LL which maps given models onto the LL gaussian boson
hamiltonian. The description in the standard -non chiral- LL should be
contrasted with that of Wen's chiral LL, a variant of the former LL, used to
describe edge states for the fractional quantum Hall effect (FQHE)\cite{wen}
where besides phonons one has fractional charge excitations: the Laughlin
quasiparticles. Actually a first step aiming at including other types of
excitations besides phonons for non-chiral LL was taken by Haldane in his
own rigorous solution of the Tomonaga-Luttinger model, which yielded neutral
collective density modes (phonons) plus charge and current excitations\cite
{haldane}. The latter excitations are generated by the zero mode part of the
boson hamiltonian and are customary in CFT\cite{ginsparg}. In the
non-interacting case Landau quasiparticles created at $\pm k_{F}$ are exact
eigenstates of these zero modes; however they are not elementary excitations
any more when interactions are turned on.

Our paper revisits those charge and current excitations for the generic
(non-chiral) LL boson hamiltonian. Due to chiral separation these charge and
current excitations split into two independent chiral components, a property
familiar from CFT; we will show that each of these chiral components must be
viewed as a \textit{composite excitation} built from two elementary
fractional objects. One of these elementary excitations will be shown to be
a Laughlin quasiparticle\cite{laughlin}; this will be established by
computing the ground state and excitations eigenfunctions for the LL boson
hamiltonian: we will prove that the ground state and one of the elementary
excitations have wavefunctions which are the 1D analogs of FQHE Laughlin
wavefunctions. The nature of the other elementary excitation depends on
whether one considers fermions or bosons: for bosons (and spins), we recover
a charge one-half object (the spinon) while for fermions we end up with a
novel object coming from the decay of the electron. We will also derive the
eigenfunctions for Wen's chiral LL for the case of filling fractions $\nu
=1/(2n+1)$.

The gaussian boson hamiltonian is just a sum of harmonic oscillators so that
the determination of the ground state and of its excitations is easy: 
\begin{equation}
H_{B}=\frac{u}{2}\int_{0}^{L}dx\ K^{-1}(\nabla \Phi )^{2}+K(\nabla \Theta
)^{2}  \label{boson}
\end{equation}
where $\Theta $ and $\Pi =\nabla \Phi $ are canonical conjugate boson
fields, $j=\frac{1}{\sqrt{\pi }}\nabla \Theta $ and $\delta \rho =-\frac{1}{%
\sqrt{\pi }}\nabla \Phi $ ($j$ and $\delta \rho $ are respectively the
particle current density and the particle density operators) and where $\Psi
_{B}=\rho ^{1/2}\exp (i\sqrt{\pi }\Theta )$ for bosons, and $\Psi _{F}=\Psi
_{B}\quad (\exp (ik_{F}r-i\sqrt{\pi }\Phi )+\exp (-ik_{F}r+i\sqrt{\pi }\Phi
))$ for fermions (we have multiplied $\Psi _{B}$ by a Jordan-Wigner phase). $%
u$ and $K$ are the usual LL parameters \cite{haldane}. The Fourier-transform
of $H_{B}$ is: 
\begin{equation}
H_{B} =\frac{u}{2}\sum_{q\neq 0}K^{-1}\Pi _{q}\Pi _{-q}+Kq^{2}\Theta
_{q}\Theta _{-q}  +\frac{\pi u}{2L}\left( \frac{\widehat{Q}^{2}}{K}+K\widehat{J}^{2}\right)
\label{qj}
\end{equation}
where $q$ is quantized as $q_{n}=2\pi n/L$, $\widehat{Q}$ and $\widehat{J}$
are the charge and the current number operators. In the ground state, $H_{B}$
reduces to a sum of harmonic oscillators (since $Q=0=J$); therefore the
ground state is just a gaussian: 
\begin{equation}
\Psi _{0}=\exp (-\frac{1}{2K}\sum_{n\neq 0}\frac{1}{\left| q_{n}\right| }\Pi
_{n}\Pi _{-n})
\end{equation}
returning to the original variables through $\Pi _{q}=\sqrt{\pi /L}\rho _{q}=%
\sqrt{\pi /L}\sum_{i}\exp (iqr_{i})$ , we find that $\Psi _{0}$ is nothing
but a Laughlin wavefunction! 
\begin{equation}
{\Psi }_{0}(\{r_{1,..,}r_{N_{0}}\})=\prod_{i<j}\mid z_{ij}{\mid }^{1/K}
\label{bosonic}
\end{equation}
(We have defined $z_{i}=\exp i2\pi r_{i}/L$ and $z_{ij}=z_{i}-z_{j}.$) This
is the correct form if we consider bosons; for fermions, we undo the
singular (Jordan-Wigner) gauge transformation converting fermions to
hard-core bosons so that ${\psi }_{0}^{F}=\prod_{i<j}$ $\left( z_{ij}\right) 
$ $\left| z_{ij}\right| ^{1/K-1}$ $\exp i\pi \frac{N-1}{L}\sum r_{i}$. This
derivation of the ground state follows exactly the same lines as that for
the bosonic Landau-Ginzburg theory for the FQHE\cite{kivelson}. We note that
eq.(\ref{bosonic}) is also the exact ground state of the Calogero-Sutherland
model\cite{cs}; the square modulus of the ground state functional of the
Thirring model was also shown to be similar\cite{fradkin}. These simple
results give a formal justification to the heuristic connection between 2D
Laughlin wavefunctions and conformal blocks of CFTs\cite{readmoore}. One
class of excited states corresponds to neutral (phonon-like) modes for which 
$Q=0$ and $J=0$. We find that the wavefunctions are simply Hermitte
polynomials: 
\begin{equation}
\left| n_{q_{1}},n_{q_{2}},...,n_{q_{p}}\right\rangle
=\prod_{s=1}^{p}H_{n_{q_{s}}}\left( \frac{\sum_{i}z_{i}^{q_{s}}}{\sqrt{%
LK\left| q_{s}\right| }}\right) \left| \Psi _{0}\right\rangle
\label{hermitte}
\end{equation}

We now focus on excitations with non-zero values for $Q$ and/or $J$. They
were first singled out by Haldane in his rigorous approach to the
Tomonaga-Luttinger model, and are standard in CFT. Such excitations are
obtained from the vertex operators (the primary operators of the CFT)\cite
{voit,ginsparg}: 
\begin{equation}
V_{Q,J}(x)=:\exp i\sqrt{\pi }\left( J\Phi (x)-Q\Theta (x)\right) :
\end{equation}
where $\left[ \widehat{Q},V_{Q,J}\right] =QV_{Q,J}$ and $\left[ \widehat{J}%
,V_{Q,J}\right] =JV_{Q,J}$. It follows from the periodicity requirement for
the fields $\Psi _{B}$ and $\Psi _{F}$ that for bosons $J$ must be an even
integer while for fermions $Q-J$ is even (both $Q$ and $J$ are integers)\cite
{haldane}. We define also $V_{Q,J}(k_{n})$ which carries a momentum $k_{n}$: 
$V_{Q,J}(k_{n})=\int dx\exp -i(k_{n}-k_{F}J)x\;V_{Q,J}(x)$. In the
non-interacting case ($K=1$) $V_{Q,J}(k_{n})$ simply describe $Q$ Landau
quasiparticles. (The bosonization formulas show indeed that the electron is $%
V_{1,\pm 1}.$) However when $K\neq 1$ Landau quasiparticles are not
elementary excitations any more (as suggested by the absence of
quasiparticle poles in the electron Green function\cite{voit}): the paradigm
set by the Heisenberg spin chain suggests that $V_{Q,J}(k_{n})|0>$ decays
into fractional excitations, much as the magnon is seen to be replaced by
two spinons. This fractionalization of the spectrum of charge and current
excitations of a LL is a direct consequence of the chiral separation of the
boson hamiltonian: $H_{B}=H_{+}+H_{-}$, and we summarize some known results
below\cite{ginsparg,voit}: 
\begin{equation}
H_{\varepsilon }=\frac{u}{4}\sum_{n\neq 0}Kq_{n}^{2}(-\varepsilon K^{-1}\Phi
_{n}+\Theta _{n})^{2}+\frac{\pi u}{LK}\widehat{Q}_{\varepsilon }^{2}
\label{chiral}
\end{equation}
where $\widehat{Q}_{\varepsilon }=\left( \widehat{Q}+\varepsilon K\widehat{J}%
\right) /2$ are chiral charges; $\Phi _{\pm }=\Phi \mp K\Theta $ are \textit{%
free} chiral fields: $\Phi _{\pm }(x,t)=\Phi _{\pm }(x\mp ut)$. It is easy
to check that $H_{+}$ and $H_{-}$ commute; that property is routinely used
in CFT.We also introduce $\Theta _{\pm }=\Theta \mp \frac{\Phi }{K}$ which
are free fields. In terms of $\Theta _{\pm }$ the charge and current
excitation operators $V_{Q,J}$ read: 
\begin{equation}
V_{Q,J}(x)=\exp -i\sqrt{\pi }Q_{+}\Theta _{+}\exp -i\sqrt{\pi }Q_{-}\Theta
_{-}
\end{equation}
When one adds $Q$ electrons (or bosons) to the system, one gets therefore
two counter-propagating parts carrying respectively a charge $Q_{+}$ and a
charge $Q_{-}$. We define chiral excitation operators: 
\begin{equation}
W_{Q_{\pm }}^{\pm }(x)=\exp -i\sqrt{\pi }Q_{\pm }\Theta _{\pm }
\end{equation}
which create a charge $Q_{\pm }$ (indeed $\left[ \widehat{Q},\exp -i\sqrt{%
\pi }Q_{\pm }\Theta _{\pm }\right] =Q_{\pm }\exp -i\sqrt{\pi }Q_{\pm }\Theta
_{\pm }$) and are easily shown to obey anyonic commutation relations with
statistics: $\mp \pi Q_{\pm }^{2}/K$ \cite{us}. The exponential in the
Fourier transform $W_{Q_{\pm }}^{\pm }(k_{n})$ can be expanded as the
product of a zero mode part and of muti-phonon processes, which shows $%
W_{Q_{\pm }}^{\pm }(k_{n})$ is an exact eigenstate of $H_{\pm }$\cite{us}.
Such an analysis in terms of chiral separation which is standard for
conformal field theorists implies that even for a spinless LL there is a
fractionalization of the electron with in general non-integral charge
excitations (since $Q_{\pm }$ is not in general an integer), a fact not
generally sufficiently appreciated in the condensed matter community.

After that summary, we now establish novel results concerning the elementary
excitations for a non-chiral LL\footnote{%
For notational convenience we will work in direct space as is customary in
CFT, the relevant exact excitations being obtained by Fourier transform.}.
For the above chiral excitations we would like to find a basis of elementary
excitations, i.e. identify objects from which all the other excitations can
be built. We must carefully distinguish between Bose and Fermi statistics
because of the constraints on $Q$ and $J$. Let us consider first bosons:
since $J$ is even we can rewrite it as $J=2n$ where $n$ is an arbitrary
integer. But then for \textit{bosons}: 
\begin{eqnarray}
\left( Q_{+},Q_{-}\right) &=&\left( \frac{Q+KJ}{2},\frac{Q-KJ}{2}\right) 
\nonumber \\
&=&Q\;\left( \frac{1}{2},\frac{1}{2}\right) +n\;\left( K,-K\right)
\label{bose-q}
\end{eqnarray}

For bosons a general excitation is therefore constructed by applying $\left(
W_{1/2}^{\pm }\right) ^{Q}\left( W_{\pm K}^{\pm }\right) ^{n}$ to $\Psi _{0}$%
, where $Q$ and $n$ are \textit{independent integers} of arbitrary sign,
which means that $W_{1/2}^{\pm }$ and $W_{\pm K}^{\pm }$ are \textit{%
elementary excitations}. (Going back to reciprocal space this means that the
exact eigenstate $W_{Q_{\pm }}^{\pm }(k_{0})$ is built from states $%
\prod_{i=1}^{Q}W_{1/2}^{\pm }(q_{i})\;\prod_{j=1}^{n}W_{K}^{\pm }(\widetilde{%
q}_{j})$ where $k_{0}=\sum_{i,j}q_{i}+\widetilde{q}_{j}.$) These \textit{two}
types of elementary excitations are generated in the following processes:
(i) adding one particle into the LL but no current ($Q=1,n=J/2=0$) results
in two (chiral) charge $1/2$ objects (with statistics $\pi /4K$\cite{us}),
moving with opposite velocities $u$, namely $W_{1/2}^{\pm }$; in spin
problems (spins are hard-core bosons) $W_{1/2}^{\pm }$ is naturally
interpreted as a $S=1/2$ spinon; (ii) creating current without addition of a
particle ($Q=0,n=1$) results in charges $K$ and $-K$ moving with opposite
velocities $u$. To identify the nature of these objects we compute for
instance $W_{K}^{+}(z)\Psi _{0}$ in first quantization; using eq.(\ref
{bosonic}) and $\exp -i\sqrt{\pi }Q\Theta (x)=\exp Q\frac{\delta }{\delta
\rho (x)}$ gives: 
\begin{eqnarray}
&&\left[ \exp Q\frac{\delta }{\delta \rho (x)}\right] \;\exp \frac{1}{2K}%
\int \int \rho (y)\ln \left| \sin \frac{\pi }{L}(y-y^{\prime })\right| \rho
(y^{\prime })  \nonumber \\
&=&\exp \frac{Q}{K}\int \rho (y)\ln \left| \sin \frac{\pi }{L}(y-x)\right|
dy\Psi _{0}  \nonumber \\
&=&C\prod_{i}\left| z_{i}-z\right| ^{Q/K}\Psi _{0}
\end{eqnarray}
($C$ is an unessential constant). Since $\exp i\sqrt{\pi }J\Phi (x)=\exp
i\pi J\int_{0}^{L}\theta (x-y)\delta \rho (y)dy$ where $\theta (x)=\frac{1}{%
i\pi }\ln [\frac{(-x)}{|x|}]$ is the step function, $\exp i\sqrt{\pi }J\Phi
(x)$ $=$ $\prod_{i}\left[ \left( z_{i}-z\right) /\left| z_{i}-z\right|
\right] ^{J}$ $\exp -iJk_{F}(2x+\sum_{i}r_{i}/N)$. Thus 
\begin{equation}
W_{K}^{+}(x)\Psi _{0}=\prod_{i}\left( z_{i}-z\right) \prod_{i<j}\mid z_{ij}{%
\mid }^{1/K}\exp -ik_{F}(2x+\frac{\sum_{i}r_{i}}{N})
\end{equation}
This leads us to identify $W_{K}^{+}$ with a Laughlin quasiparticle; the
charge deduced from a plasma analogy is of course $K$ in agreement with the
above considerations. A simple argument may allow to appreciate the full
parallel between the way Laughlin quasiparticles are generated in the FQHE
and in a (non-chiral) LL. Laughlin showed that insertion of a flux quantum
creates a fractionally charged Laughlin quasihole in the bulk\cite{laughlin}%
; by charge conservation the opposite charge is created at the edge\cite{L81}%
. We now show that threading flux in a LL ring also leads to the creation of
a Laughlin quasiparticle-quasihole pair. Insertion of flux $\phi $ leads to
the replacement $\frac{\pi u}{2L}K\widehat{J}^{2}\longrightarrow \frac{\pi u%
}{2L}K\left( \widehat{J}+2\frac{\phi }{\phi _{0}}\right) ^{2}$ in eq.(\ref
{qj})\cite{schulz}. When $\phi =\phi _{0}$ the energy is minimized for $J=-2$
and the state is $V_{0,-2}$ $=$ $\exp -i2\sqrt{\pi }\Phi .$ This is a ($%
Q=0,J=-2)$ process which by the previous analysis corresponds to the
creation of $W_{K}^{+}$ and $W_{-K}^{-}$ a Laughlin quasiparticle-quasihole
pair. This is entirely analogous to Laughlin's thought experiment. We note
that Fisher and Glazman had previously argued for the existence of charge $K$
Laughlin quasiparticles in a LL, but their argument was heuristic, relying
on a study of backscattering by a barrier in a LL\cite{fisher}, while the
existence of charge $K$ elementary excitations was derived in this paper
from first principles. Since fractional charge does not mean Laughlin
quasiparticle (charge 1/3 solitons were known from Su and Schrieffer's work
on charge density waves\cite{su}), it is also necessary to prove that these
charge $K$ objects actually are genuine Laughlin quasiparticles: this was
established in our computation of the wavefunction for $W_{K}^{+}$.

Let us illustrate these results on the specific example of the anisotropic
Heisenberg $XXZ$ spin chain. It is crucial to choose the right selection
rules for $Q$ and $J$: either bosonic or fermionic; here the correct ones
are the bosonic ones because $S=1/2$ spins are bosons (spins are \textit{not}
Jordan-Wigner fermions due to the string factor). Then consider spin
transitions: $\Delta S=1$ (i.e. $Q=1,J=0$); when we vary the anisotropy we
observe a continuum of excitations wich can be identified with spinons.This
is recovered easily: indeed for $Q=1,J=0$ the previous results show that we
generate $W_{1/2}^{+}$ and $W_{1/2}^{-}$. (Note that at the isotropic point
for which $K=1/2$ our spin half $W_{1/2}^{+}$ has a semionic statistics $\pi
/2$ but in general its exchange statistics is $\pi /4K$\cite{us}$.$ The
spinon operator and its exclusion statistics have been considered in the $%
SU(2)$ symmetric case -the isotropic point- in \cite{spinon}.) The spinon
operator $W_{1/2}^{+}$ upon acting on the ground state yields $W_{1/2}^{\pm
}(z)\Psi _{0}=\prod_{i}\left( z_{i}-z\right) ^{1/2K}$ $\prod_{i<j}$ $\mid
z_{ij}{\mid }^{1/K}$ $\exp -ik_{F}/2K(2x+\sum_{i}r_{i}/N)$. Besides spinons
we have the novel result that there should also be spin $K$ Laughlin
quasiparticles generated by spin currents. Imposing a twist in the boundary
conditions will give rise to a Laughlin quasiparticle- quasihole pair of
charge $\pm K$. We note that for the special case $K=1/2$ (the isotropic
Heisenberg chain, or boson self-dual point) elementary excitations consist
solely of spinons and antispinons.

We now turn to fermions and we may write $Q-J=2n$. Then for \textit{fermions}%
: 
\begin{equation}
\left( Q_{+},Q_{-}\right) =Q\;\left( \frac{1+K}{2},\frac{1-K}{2}\right)
-n\;\left( K,-K\right)  \label{first}
\end{equation}

Once again we may take ($Q=1,n=0$); this corresponds to chiral charges $%
\frac{1-K}{2}$ and $\frac{1+K}{2}$ moving at velocities $-u$ and $u$
respectively. The total current is $uK$ (in units of $k_{F}$): indeed $n=0$
implies $J=1$, highlighting the fact that for fermions elementary
excitations may mix charge and current states. Another generator is obtained
with ($Q=0,n=1$): it is a pure current state made out of counterpropagating
Laughlin quasiparticle ($+K$) and anti-particle ($-K$). Note that in the
non-interacting case ($K=1$) these quasiparticles reduce to bare particles
populating states near $\pm k_{F}$. The general excitation is again built by
creating an integer number of times $W_{\frac{1\pm K}{2}}^{\pm }$ and/or $%
W_{\pm K}^{\pm }$ states which precisely means that we have identified a set
of elementary excitations for the fermionic LL.

The elementary excitations we have derived form a basis from which all
charge and current excitations are obtained; by no means are they the only
choice of basis: other bases of elementary excitations are obtained by
considering base changes matrices with integer entries whose inverses are
also integer-valued, which ensures that all excitations are integral linear
combinations of the elementary excitations. For instance for fermions
another basis is $W_{\frac{1\pm K}{2}}^{\pm }$and $W_{1}^{\pm }$: 
\begin{equation}
\left( Q_{+},Q_{-}\right) =J\;\left( \frac{1+K}{2},\frac{1-K}{2}\right)
+n\;\left( 1,1\right)  \label{dual}
\end{equation}
it is actually a basis dual to the previous one (obtained under exchanges of 
$K\longleftrightarrow 1/K$ and $\Phi \longleftrightarrow \Theta $).

Turning to Wen's chiral LL for the filling fraction $\nu =1/(2n+1)$ using
Wen's modified expression for the electron operator $\Psi =\exp
i(2n+1)\varphi $ and $H=2\pi (2n+1)\sum_{k>0}\rho _{k}\rho _{-k}$ (with $%
\rho =\frac{1}{2\pi }\nabla \varphi $) one finds the ground state ${\psi }%
_{0}=\prod_{i<j}\left( z_{ij}\right) ^{2n+1}\exp -ik_{F}\sum r_{j}$ which up
to a current term is just the 1D restriction of Laughlin bulk wavefunction;
similarly the fractional charge excitation $\exp i\varphi /(2n+1)$ yields
the wavefunction $\prod_{i}(z_{i}-z)\prod_{i<j}(z_{i}-z_{j})^{2n+1}$. To our
knowledge this is the first time the wavefunctions of Wen's chiral LL are
directly computed from the hamiltonian: while a heuristic connection had
already been made between Laughlin's bulk wavefunctions and 1D edge theories
(2D bulk wavefunctions as conformal blocks of 1+1D CFTs), we believe our
results make the relation quite transparent.

The previous discussion can be generalized to the LL with spin; one should
consider multicomponent Laughlin wavefunctions ${\psi }_{0}(\{r_{i},\sigma
_{i}\})=\prod_{i<j}\mid z_{ij}{\mid }^{g_{\sigma _{i},\sigma _{j}}}$. The
discussion of charge and current excitations in a spinful LL then parallels
the spinless case; in the case of spin-charge separation one finds a basis
of four elementary excitations among which a charge one holon and a spin
half spinon, as well as ''Laughlin holons'' carrying a charge $K_{c}$ and
''Laughlin spinons'' with spin $K_{s}/2$.

Experimental observation of the various fractional excitations considered in
this paper is of course an important issue in the extent that there are
several candidates for a realization of the LL: quasi-1D organics, quantum
wires, gapless spin chains, quantum Hall edge states\cite{voit}; an
intriguing possiblity might also be chiral edges of compressible phases of
the 2D electron gas in between Hall plateaux\cite{grayson}. Relevant probes
could be as in the FQHE shot noise experiments or resonant tunneling\cite
{exp}. We summarize the novel results obtained in the present paper: we have
identified the elementary excitations of the LL, which are fractional
objects (carrying fractional charges); by computing their wavefunctions we
have found that one of them is identified as a Laughlin quasiparticle and is
generated by current excitations in a LL. This property is a natural
consequence of the fact that the ground state of the LL boson hamiltonian is
nothing but a Laughlin wavefunction. We have also computed the
eigenfunctions for Wen's chiral LL at filling fractions $\nu =1/(2n+1)$
which are just the one dimensional restrictions of the 2D bulk states. The
authors thank B.Jancovici, D.Bazzali for interesting discussions, as well as
the Orsay theory group, and especially T. Giamarchi and Heinz Schulz for
their comments.

\end{document}